\documentclass[twocolumn]{aastex631}
\usepackage{amsmath}

\shorttitle{Spatially-Resolved Recent Star Formation History in NGC 6946}
\shortauthors{Tran et al.}
\graphicspath{{./}{figures/}}

\begin{document}

\title{Spatially-Resolved Recent Star Formation History in NGC 6946}

\author[0000-0002-6440-1087]{Debby Tran}
\affiliation{Department of Astronomy, Box 351580, University of Washington, Seattle, WA 98195, USA}
\author[0000-0002-7502-0597]{Benjamin Williams}
\affiliation{Department of Astronomy, Box 351580, University of Washington, Seattle, WA 98195, USA}
\author[0000-0003-2184-1581]{Emily Levesque}
\affiliation{Department of Astronomy, Box 351580, University of Washington, Seattle, WA 98195, USA}
\author[0000-0002-0786-7307]{Margaret Lazzarini}
\affiliation{Division of Physics, Mathematics, and Astronomy, California Institute of Technology, 1200 E California Boulevard, Pasadena, CA 91125, USA}
\author[0000-0002-1264-2006]{Julianne Dalcanton}
\affiliation{Center for Computational Astrophysics, Flatiron Institute, 162 Fifth Avenue, New York, NY 10010, USA}
\author[0000-0001-8416-4093]{Andrew Dolphin}
\affiliation{Raytheon Technologies, 1151 E. Hermans Road, Tucson, AZ 85756, USA}
\affiliation{University of Arizona, Steward Observatory, 933 N. Cherry Avenue, Tucson, AZ 85721, USA}
\author[0000-0001-5530-2872]{Brad Koplitz}
\affiliation{School of Earth \& Space Exploration, Arizona State University, 781 Terrace Mall, Tempe, AZ 85287, USA}
\author[0000-0003-2599-7524]{Adam Smercina}
\affiliation{Department of Astronomy, Box 351580, University of Washington, Seattle, WA 98195, USA}
\author[0000-0003-4122-7749]{O. Grace Telford}
\affiliation{Department of Physics and Astronomy, Rutgers University, 136 Frelinghuysen Road, Piscataway, NJ 08854, USA}

\begin{abstract}
The nearby face-on star forming spiral galaxy NGC 6946 is known as the Fireworks Galaxy due to its hosting an unusually large number of supernova. We analyze its resolved near-ultraviolet (NUV) stellar photometry measured from images taken with the Hubble Space Telescope's (HST) Wide Field Camera 3 (WFC3) with F275W and F336W filters. We model the color-magnitude diagrams (CMD) of the UV photometry to derive the spatially-resolved star formation history (SFH) of NGC 6946 over the last 25 Myr. From this analysis, we produce maps of the spatial distribution of young stellar populations and measure the total recent star formation rate (SFR) of nearly the entire young stellar disk. We find the global SFR(age$\leq$25 Myr)=$13.17 \substack{+0.91 \\-0.79} M_\odot/\rm yr$. Over this period, the SFR is initially very high ($23.39\substack{+2.43\\-2.11} M_\odot/\rm yr$ between 16-25 Myr ago), then monotonically decreases to a recent SFR of $5.31\substack{+0.19\\-0.17} M_\odot/\rm yr$ in the last 10 Myr. This decrease in global star formation rate over the last 25 Myr is consistent with measurements made with other SFR indicators. We discuss in detail two of the most active regions of the galaxy, which we find are responsible for 3\% and 5\% of the total star formation over the past 6.3 Myr. 
\end{abstract}

\keywords{Star formation(1569); Galaxy evolution(594); Stellar populations(1622); Disk galaxies(391)}

\section{Introduction} \label{sec:intro}

Star formation rate (SFR) is one of the defining characteristics in determining the current evolutionary state of a galaxy. The SFR strongly affects evolution through metal production \citep{Larson1974}, gas consumption \citep{Kennicutt1983,Chiappini1997}, cold gas content \citep{Kauffmann2000}, and feedback in the galaxy \citep{deRossi2009}. Thus, SFR is a key property in tests of galaxy evolution models \citep{Tinsley1980,Kennicutt1998,Kauffmann2000}. Because of its significance, many methods of measuring star formation rate with observational data have been developed, such as measuring UV emission from young ($\lesssim10$ Myr) massive stars \citep{Kennicutt1998,Kennicutt2012}, H$\alpha$ emission from the youngest ($\lesssim5$ Myr) massive stars \citep{Kennicutt1983,Shivaei2015}, and estimating SFR from the rate of core-collapse supernova (ccSN) \citep{Eldridge2019}, which probes SFR at timescales at which stars supernova (30-100 Myr ago). These methods probe a range of timescales, making direct comparisons between SFR measured with different indicators challenging. Star formation histories (SFHs) avoid this problem by providing the SFR over time, allowing us to compare and calibrate SFR measurements obtained with different methods, while revealing how galaxies change over time.

Even more powerful are spatially-resolved star formation histories, which provide both temporal and spatial information. With these, we can trace local mechanisms that could be triggering star formation. For nearby galaxies with resolved stellar photometry, we can construct and fit observed color-magnitude diagrams (CMDs) to infer a star formation history for a specific region, assuming a specific initial mass function (IMF), stellar evolutionary model, binary fraction, and distribution of dust. By tiling together SFHs from multiple regions, we can construct a spatially-resolved star formation history for a galaxy. This kind of work has been done in the Small and Large Magellenic Clouds \citep{Harris2004,Harris2009}, M31 \citep{Williams2003,Lewis2015}, M33 \citep{Lazzarini2022}, and M81 \citep{Choi2015}.

In this paper, we apply this technique to NGC 6946, which has been widely studied due to its active star formation (\cite{Schinnerer2007} classify it as a circumnuclear starburst) and the high frequency of supernovae in the past century \citep{Sauty1998,Meier2004,Schinnerer2007,Murphy2011,Kennicutt2011,Botticella2012,Tsai2013,Gorski2018,Eldridge2019,Eibensteiner2022}. Among these studies, there have been inconsistent measurements of the galaxy's global star formation rate, ranging widely from 3-12 $M_\odot/\rm yr$, due to the diverse methods of measuring star formation rate and wide range of different distances used. Throughout this paper, we use a distance of 7.83 $\pm$ 0.29 Mpc \citep{Murphy2018} and an inclination of 32.8$\arcdeg$ \citep{deBlok2008}. \cite{Murphy2011} has explored the accuracy of these various diagnostics for a sample of regions in NGC 6946, finding discrepancies of up to factors of 5. To better constrain the recent star formation history across the entire galaxy, we have carried out a NUV HST survey to obtain photometry of the young massive stellar population of NGC 6946. This dataset provides the most detailed and complete probe to date of the global, localized, and episodic star formation in NGC 6946.

\begin{figure*}[h]
    \centering
    \includegraphics[width=\textwidth]{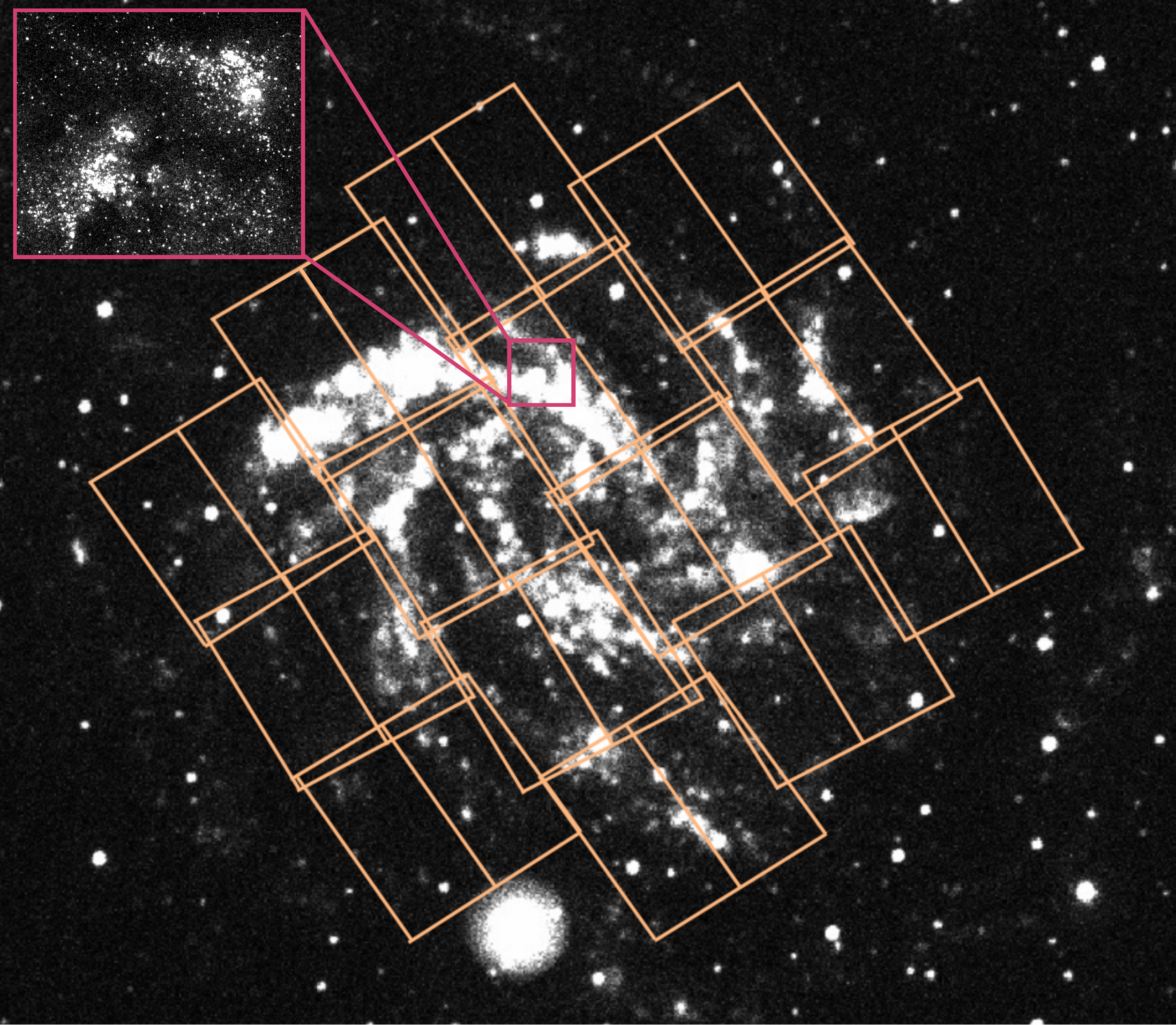}
    \caption{Yellow boxes showing HST footprint of the 14 pointings of our data overlaid on a GALEX NUV image from the GALEX Nearby Galaxies Survey (NGS; \cite{GildePaz2007} and references therein). In the magenta thumbnail is an example of what the resolution of the HST data in F336W to illustrate the high spatial resolution that allowed us to get spatially-resolved stellar photometry.}
    \label{fig:footprint}
\end{figure*}

In Section \ref{sec:data}, we present the HST observations, alignment of the data, photometry, artificial star tests for measuring photometric uncertainties, gridding schema, and method for measuring star formation rates.  In Section \ref{sec:results}, we present the recent star formation rates, the reliability of the SFRs at the youngest and oldest time bins, total stellar mass formed over the past 25 Myr, and foreground and differential extinction of each cell. In Section \ref{sec:discussion}, we discuss two highly star-forming regions of interest, the decline in global star formation rate, and the correlation between stellar density and age. In Section \ref{sec:conclusions}, we summarize our methods and findings.

\section{Observations and Data Analysis} \label{sec:data}
Observations for this program (GO-15877; PI \cite{Levesque2019prop}) were obtained between May 11 2020 and November 21 2021 using HST's WFC3 Ultraviolet- (UVIS) channel in filters F275W and F336W. Details of the observations are found in Table \ref{tab:obs}. NGC 6946 was imaged in a 4x4 grid excluding the northernmost and southernmost regions (Figure \ref{fig:footprint}). This covers all of the UV-bright regions and the locations of observed core collapse supernovae. Each neighboring pointing overlaps at the edges to ensure there are no gaps in the catalog due to poorer photometric quality at the edges. Each pointing in both filters was dithered with small offsets to control for hot pixels and cosmic rays. Unfortunately, even with the careful selection of observing strategy, there are two small gaps of $10\arcsec\times1\arcmin$ and $30\arcsec\times1\arcmin$ approximately centered at 20:34:39.40 +60:06:80 and 20:34:25.00 +60:08:80, respectively. These gaps are due to adjusting the rotation of two pointings to obtain a sufficient number of guide stars. Upon comparison with existing optical data, there do not appear to be dense star clusters in these two gaps. 

\begin{deluxetable*}{cccccccc}
\tablenum{1}
\tablecaption{Details of Fields Observed \label{tab:obs}}
\tablewidth{0pt}
\tablehead{
\colhead{Field Name} & \colhead{R.A.} & \colhead{Dec} & \colhead{Filters} &
\colhead{Exposure Time} & \colhead{Number of} & \colhead{Date} & \colhead{Roll Angle}\\
\colhead{} & \colhead{(hh:mm:ss.sss)} & \colhead{(\arcdeg\arcdeg:\arcmin\arcmin:\arcsec\arcsec.\arcsec\arcsec)} & \colhead{} &
\colhead{(s)} & \colhead{Exposures} & \colhead{(YYYYMMDD)} & \colhead{(PA\_V3)}
}
\startdata
NGC6946-2 & 20:35:06.007 & +60:12:48.93 & F275W & 1432 & 2 & 20201106 & 257.0 \\
NGC6946-2 & 20:35:06.007 & +60:12:48.93 & F275W & 1414 & 2 & 20201106 & 257.0 \\
NGC6946-3 & 20:34:38.218 & +60:12:59.17 & F275W & 1432 & 2 & 20201109 & 256.5 \\
NGC6946-3 & 20:34:38.218 & +60:12:59.17 & F275W & 1414 & 2 & 20201109 & 256.5 \\
NGC6946-4 & 20:35:19.114 & +60:10:49.51 & F275W & 1432 & 2 & 20201110  & 255.7 \\
NGC6946-4 & 20:35:19.114 & +60:10:49.51 & F275W & 1414 & 2 & 20201110 & 255.7 \\
NGC6946-5 & 20:34:51.355 & +60:10:59.92 & F275W & 1432 & 2 & 20201103 & 257.0 \\
NGC6946-5 & 20:34:51.355 & +60:10:59.92 & F275W & 1414 & 2 & 20201103 & 257.0 \\
NGC6946-6 & 20:34:23.591 & +60:11:09.97 & F275W & 1432 & 2 & 20201103 & 257.0  \\
NGC6946-6 & 20:34:23.591 & +60:11:09.97 & F275W & 1414 & 2 & 20201103 & 257.0 \\
NGC6946-7 & 20:35:32.194 & +60:08:50.01 & F275W & 1432 & 2 & 20201112 & 257.0\\
NGC6946-7 & 20:35:32.194 & +60:08:50.01 & F275W & 1414 & 2 & 20201111 & 257.0\\
NGC6946-8 & 20:35:04.464 & +60:09:00.59 & F275W & 1432 & 2 & 20201112 & 253.5 \\
NGC6946-8 & 20:35:04.464 & +60:09:00.59 & F275W & 1414 & 2 & 20201112 & 253.5 \\
NGC6946-9 & 20:34:36.730 & +60:09:10.81 & F275W & 1371 & 2 & 20210504 & 75.5\\
NGC6946-9 & 20:34:36.730 & +60:09:10.81 & F275W & 1390 & 2 & 20210504 & 75.5\\
NGC6946-10 & 20:34:09.061 & +60:09:21.20 & F275W & 1432 & 2 & 20200515 & 73.0\\
NGC6946-10 & 20:34:09.061 & +60:09:21.20 & F275W & 1414 & 2 & 20200515 & 73.0\\
NGC6946-11 & 20:35:17.548 & +60:07:01.19 & F275W & 1432 & 2 & 20201112  & 253.0\\
NGC6946-11 & 20:35:17.548 & +60:07:01.19 & F275W & 1414 & 2 & 20201112 & 253.0 \\
NGC6946-12 & 20:34:49.842 & +60:07:11.57 & F275W & 1432 & 2 & 20201113 &253.0\\
NGC6946-12 & 20:34:49.842 & +60:07:11.57 & F275W & 1414 & 2 & 20201113 & 253.0\\
NGC6946-13 & 20:34:22.203 & +60:07:22.13 & F275W & 1432 & 2 & 20200511 & 73.0\\
NGC6946-13 & 20:34:22.203 & +60:07:22.13 & F275W & 1414 & 2 & 20200511 & 73.0\\
NGC6946-14 & 20:35:02.928 & +60:05:12.26 & F275W & 1432 & 2 & 20201107 & 257.0\\
NGC6946-14 & 20:35:02.928 & +60:05:12.26 & F275W & 1414 & 2 & 20201107 & 257.0\\
NGC6946-15 & 20:34:35.318 & +60:05:22.98 & F275W & 1362 & 2 & 20211114 & 257.0\\
NGC6946-15 & 20:34:35.318 & +60:05:22.98 & F275W & 1361 & 2 & 20211115 & 257.0\\
NGC6946-2 & 20:35:06.007 & +60:12:48.93 & F336W & 880 & 3 & 20201106  & 257.0\\
NGC6946-3 & 20:34:38.218 & +60:12:59.17 & F336W & 880 & 3 & 20201109 & 256.5 \\
NGC6946-4 & 20:35:19.114 & +60:10:49.51 & F336W & 880 & 3 & 20201110 & 255.7\\
NGC6946-5 & 20:34:51.355 & +60:10:59.92 & F336W & 880 & 3 & 20201103  & 257.0\\
NGC6946-6 & 20:34:23.591 & +60:11:09.97 & F336W & 880 & 3 & 20201103 & 257.0 \\
NGC6946-7 & 20:35:32.194 & +60:08:50.01 & F336W& 880 & 3 & 20201111 & 257.0 \\
NGC6946-8 & 20:35:04.464 & +60:09:00.59 & F336W & 880 & 3 & 20201112 & 253.5\\
NGC6946-9 & 20:34:36.730 & +60:09:10.81 & F336W & 865 & 3 & 20210504 &75.5\\
NGC6946-10 & 20:34:09.061 & +60:09:21.20 & F336W & 880 & 3 & 20200515 & 73.0\\
NGC6946-11 & 20:35:17.548 & +60:07:01.19 & F336W & 880 & 3 & 20201112 & 253.0\\
NGC6946-12 & 20:34:49.842 & +60:07:11.57 & F336W & 880 & 3 & 20201113 &253.0\\
NGC6946-13 & 20:34:22.203 & +60:07:22.13 & F336W & 880 & 3 & 20200511& 73.0 \\
NGC6946-14 & 20:35:02.928 & +60:05:12.26 & F336W & 880 & 3 & 20201106 &257.0\\
NGC6946-15 & 20:34:35.318 & +60:05:22.98 & F336W & 870 & 3 & 20211114 & 257.0 \\
\enddata
\tablecomments{Config Mode - WFC3/UVIS Imaging}
\end{deluxetable*}

\subsection{Source Detection and Photometry} \label{ssec:photometry}
HST WFC3 NUV photometry were measured using DOLPHOT \citep{Dolphin2000,Dolphin2016}, a stellar photometry package using point spread function (PSF) fitting, described in detail in \cite{Williams2014}. We generated separate catalogs for each of the 14 overlapping pointings using the same DOLPHOT parameters as in \cite{Williams2014}. We then combined all of the measurements into a single catalog, described in Section \ref{ssec:astrometry}. In this catalog, we identified sources as reliable, high quality photometry using the metrics of sharpness$^2 < 0.2$; crowding $<$ 0.7; signal-to-noise ratio (SNR) $>$ 4 in both F275W and F336W; and F275W-F336W color $<$ -1.3, as anything blueward of this color is unphysical for young massive stars, see Figure \ref{fig:cmd} for comparison with Padova isochrones. For the analysis in the paper, we used $\sim$ 81,000 sources that passed the aforementioned quality cuts. The brightest single stars in the Padova log(age)=6.6 isochrone \citep{Marigo2008}, the youngest age we could fit, had a F336W magnitude of 20, so sources with magnitudes brighter would be likely blends. These likely blends, which are noted in the catalog, were included in our analysis as we were interested in the high crowding regions. The impact of including the blends is further discussed in Section \ref{ssec:young-reliability}.

\subsection{Astrometry and Foreground Stars} \label{ssec:astrometry}
Using the high quality photometry, we cross-match our stellar catalog to \textit{Gaia} Data Release 2 (\textit{Gaia} DR2; \cite{Gaia2016}; \cite{Gaia2018}). We shifted each frame by the median of the residuals of the sources matched between our catalog and \textit{Gaia} DR2. The residuals have mean magnitudes on milliarcsecond scales in both right ascension and declination, which is roughly a hundred times smaller than a WFC3 UVIS pixel. The values of the overlapping sources were then averaged. 

After finding \textit{Gaia} matches, we removed likely foreground stars from our analysis. First, we utilized the matched \textit{Gaia} sources to remove anything with a measured proper motion, as it is a likely foreground star. Then, we applied the following F275W-F336W color cuts to remove brighter, redder sources that are likely foreground stars: F336W $<$ 21, color $>$ 0.7; F336W $<$ 22, color $>$ 1; F336W $<$ 22.5, color $>$ 2; F336W $<$ 23, color $>$ 2.5. The photometric catalog used in this paper can be found as a High Level Science Product in MAST (the Mikulski Archive for Space Telescopes) via doi:\dataset[10.17909/gveq-8820]{http://dx.doi.org/10.17909/gveq-8820}. 

\subsection{Spatial Mapping} \label{ssec:grids}
\begin{figure*}
    \centering
    \includegraphics[width=1\textwidth]{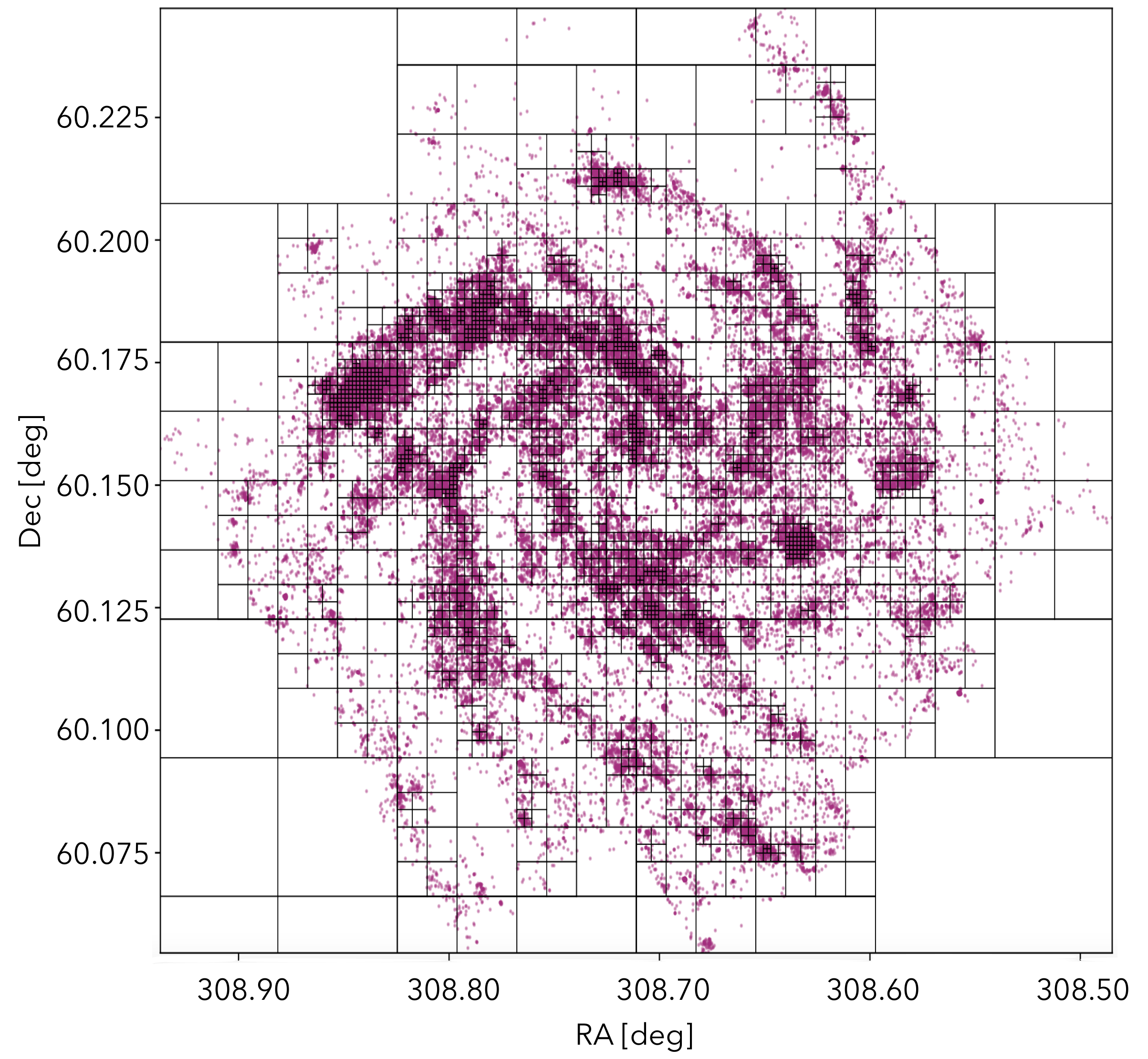}
    \caption{Grid pattern determined via quadtree with the positions of the good quality stars in pink. The quadtree algorithm works by iteratively subdividing regions that exceed a set number of stars into four equal squares until it hits a minimum cell size of 3"x3" (100 pc x 100 pc), which is roughly the size of clusters. For more details on our implementation of the quadtree algorithm, see Section \ref{ssec:grids}}
    \label{fig:grid}
\end{figure*}

To recover the spatially-resolved SFH, we first divide our full photometric catalog into a custom grid pattern (Figure \ref{fig:grid}), allowing us to recover the SFH in each grid cell independently. We choose a grid pattern such that the size of the cell is based on the stellar density of the cell, which helps equalize the number of stars per cell. This gridding schema ensures the denser regions are divided into finer spatial bins, or cells, taking advantage of the large number of stars available for age constraints. Conversely, the less dense regions are divided into coarser cells to ensure each cell has enough stars to measure reliable ages (see Section \ref{ssec:old-reliability} for details).

To generate the cell vertices, we implemented a quadtree algorithm, which operates as follows. First, it counts the number of stars in the specified region. If the number of stars in the region is higher than a certain threshold (100 stars in this study), then it will subdivide into four equal parts. This iterates until it hits a minimum cell size, which is roughly 3$\arcsec\times3\arcsec$ ($\sim 100 \rm pc\times100 \rm pc$), chosen because that is the approximate size of clusters in NGC 6946. 

\subsection{Artificial Star Tests and Completeness} \label{ssec:ASTs}
We use artificial star tests (ASTs) to measure the effects of noise, crowding, and bias on the photometry. We injected artificial stars into regions of the galaxy of different stellar densities to assess how well artificial stars of different colors and magnitudes are recovered as a function of stellar density. The input and recovered colors and magnitudes are included as a parameter in the derivation of the SFHs (Section \ref{ssec:derivation}) to account for these biases.

\begin{figure*}[h]
    \centering
    \includegraphics[width=1\textwidth]{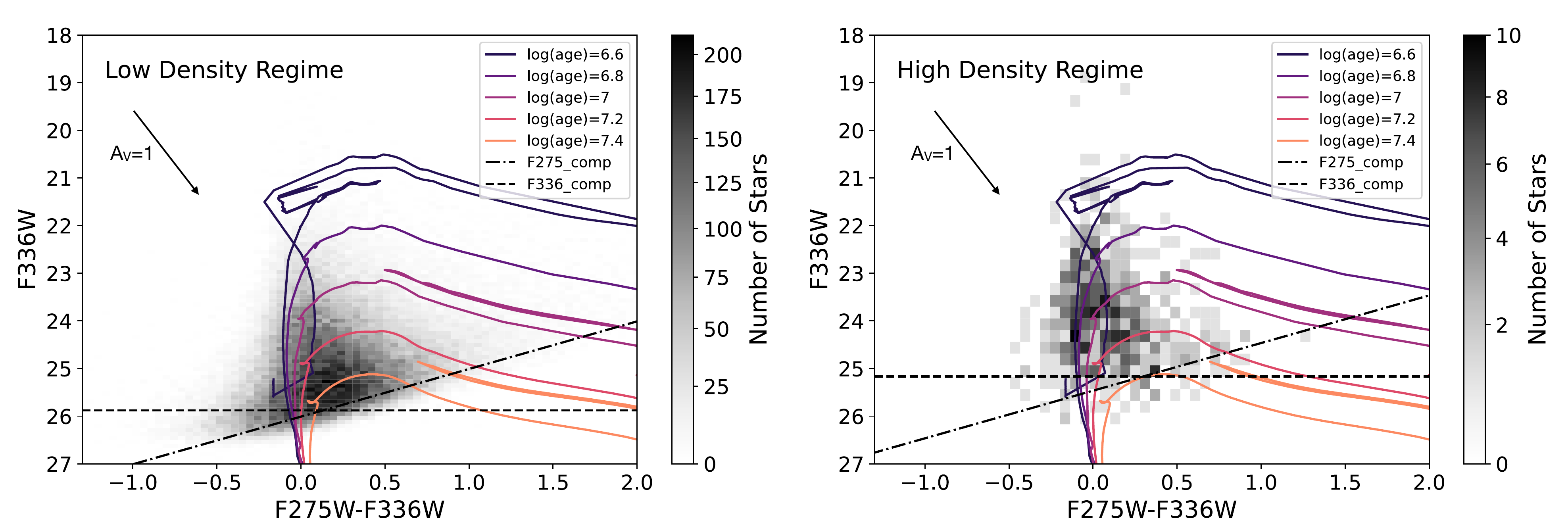}
    \caption{Left: The binned observed color-magnitude diagram of all the stars in density bins between 0-11.5 stars/arcsec$^2$. Right: The binned observed CMD of all stars in density bins between 12-17 stars/arcsec$^2$. On both CMDs, isochrones with log(age)=6.6 to log(age)=7.4, the F275W and F336W 50\% completeness, and an A$_V$=1 vector are overplotted.}
    \label{fig:cmd}
\end{figure*}

Because the impacts of crowding are largely density-dependent, we must ensure that our ASTs are fully sampling the wide range in the environment. For each cell, we calculate a stellar density by taking the number of stars that pass the quality cut described in Section \ref{ssec:photometry} above 25 mag in the F336W filter in the cell and dividing it by size of cell in arcsec$^2$. We attempted several ways of binning the cells by density, but ultimately separated them into a low density regime (cells with densities less than 11.5 stars/arcsec$^2$) and a high density regime (cells with densities greater than 11.5 stars/arcsec$^2$). We illustrate the differences in the depth of the observed data in the low and high density regimes in Figure \ref{fig:cmd}.

We further bin these cells by density to generate at least 20,000 artificial stars per density bin to ensure that we have a sufficiently fine grid of artificial stars of different colors and magnitudes. These artificial stars were then run through DOLPHOT and flagged as recovered or unrecovered. Artificial stars are defined as recovered if they pass the same quality cuts we apply to our dataset, described in Section \ref{ssec:photometry}. For each density bin, we divided recovered and unrecovered stars into bins of width of 0.2 magnitude. We then convolved the ratio of recovered to unrecovered stars with a boxcar function to smooth this ratio and interpolated the magnitudes at which 50\%, of the stars are recovered, or simply the 50\% completeness limits. The F275W and F336W 50\% completeness limits define the magnitude ranges that we fit with models to obtain star formation history measurements, as described in detail in Section \ref{ssec:derivation}. 

\begin{figure*}[h]
    \centering
    \includegraphics[width=0.7\textwidth]{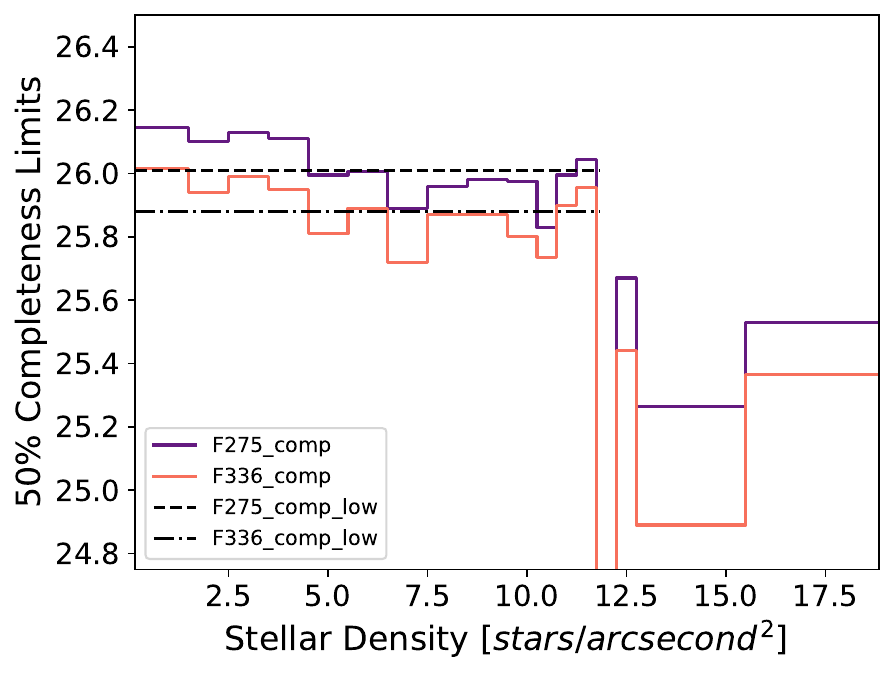}
    \caption{50\% completeness limits in both F336W and F275W plotted as a function of stellar density. The mean completeness limit of the low density regime is overplotted in the dashed lines. There are no stars with densities between 11.5-12 and are thus not plotted.}
    \label{fig:completeness}
\end{figure*}

We took the mean of the 50\% completeness limits in the low density regime to smooth over some of the stochasticity ($\pm$ 0.1 mag variation in both filters) and determined a mean 50\% completeness of 26.02 and 25.88 in the F275W and F336W filters, respectively (Figure \ref{fig:completeness}). With so few of the cells in the high density regime, it was computationally feasible to run these artificial star tests for those individual high density cells to determine the cell's individual completeness limit.

\subsection{Derivation of the SFHs} \label{ssec:derivation}
We used the CMD-fitting code, MATCH \citep{Dolphin2002}, to derive the star formation history of each cell. For each cell, MATCH creates Hess-diagrams or binned CMDs of stars in the cell. MATCH then takes user-defined ranges in age, metallicity, distance, extinction, IMF, and binary fraction to create individual synthetic CMDs for each possible combination of parameters. The individual CMDs generated from given parameters are linearly combined to form composite CMDs, which are compared to the observed CMDs. The best-fit composite synthetic CMDs are then used to infer what ages and metallicities make up the observed cell and its resulting star formation history. 

We choose a Kroupa IMF \citep{Kroupa2001}, binary fraction of 0.35, and the Padova stellar evolutionary models \citep{Marigo2008,Girardi2010}. We use the distance of 7.83 $\pm$ 0.29 Mpc to be consistent with \cite{Murphy2018}, who used the same CMD-fitting technique. Because of the short timescale, we fix the metalliciities to be between log(Z)= -0.5 to 0.1 and fix the most recent time bins to have near solar metallicities. The youngest age we could fit was log(age)=6.6. As seen in Figure \ref{fig:cmd}, the NUV data barely graze the log(age)=7.5 isochrone for the low density bin and the log(age)=7.4 isochrone for the high density bin. However, for completeness, we fit up to ages of log(age)=7.5. For a more detailed discussion on the reliability of the age range fit, see Sections \ref{ssec:young-reliability} and \ref{ssec:old-reliability}. A summary of these parameters is provided in Table \ref{tab:params}. 

\begin{deluxetable}{cc}[h!]
\tablenum{2}
\tablecaption{MATCH Fitting Parameters \label{tab:params}}
\tablewidth{0pt}
\tablehead{
\colhead{Parameter} & \colhead{Values}
}
\startdata
IMF Model & Kroupa\\
Evolutionary Models & Padua2006 \\
Distance & 7.83 $\pm$ 0.29 Mpc\\
Distance Modulus & 29.47 ± 0.079 \\
A$_V$ & 0.8-2.2, steps of 0.1 \\
log(Z) & -0.5 - 0.1, steps of 0.1 \\
Binary fraction & 0.35 \\
F336W step size & 0.1 \\
F275W-F336W step size & 0.05 \\
CMD smoothing param & 3\\
F275-F336W & -1.3-3.3 \\
Ages (log(yr)) &  6.6-7.4 for $\rho<11.5$\\
& 6.6-7.5 for $\rho \geq 11.5$\\
Age step size & 0.1\\
\enddata
\tablecomments{$\rho$ = stellar density in stars/arcsec$^2$}
\end{deluxetable}

First, we determine the best fit values of foreground extinction (A$_V$) and differential, or circumstellar, extinction (dA$_V$) of each cell by running SFH calculations over a coarse grid of A$_V$ and dA$_V$ with A$_V$ between 0.8 to 2.2 with the parameters in Table \ref{tab:params}. Finding the highest likelihood value of A$_V$ and dA$_V$, we then redo the same SFH calculations over a finer grid of values in 0.05 increments to find the best fit A$_V$ and dA$_V$, described in detail in \cite{Lewis2015}. Second, we adopt the best fit A$_V$ and dA$_V$ and rerun the SFH calculations to determine the best fit star formation rate and metallicity per time bin in each cell. Third, we measure the uncertainties of the star formation rates and metallicities by running a hybrid Monte Carlo algorithm, described in detail in Section \ref{sssec:uncertainties}. We then combine the SFH of each cell to create a spatially-resolved map of NGC 6946's star formation history, presented in Section \ref{ssec:SFR_maps}.

\subsubsection{Uncertainties} \label{sssec:uncertainties}
There are a few systematic uncertainties associated with this analysis. First, the choice of binary fraction could have a systematic impact our results. For consistency with other work deriving the recent SFH of galaxies \citep{Lewis2015,Lazzarini2022}, we adopt a binary fraction of 0.35, knowing that massive stars have a binary fraction greater than 0.7 \citep{Sana2012}. \cite{Lewis2015} showed that uncertainties introduced by choice of binary fraction are small compared to the uncertainties due to dust. Second, choice of stellar evolutionary model has a systematic impact on our results. \cite{Lazzarini2022} showed that the SFH measured using the Padova versus MIST models differed at ages less than 20 Myr. However, results from individual cells fit with both models agreed within less than 1\%. Third, choice of IMF could impact our results. For consistency, we used the Kroupa IMF, which has been widely used for measuring star formation rate in NGC 6946.

To characterize the random uncertainties, we used a hybrid Monte Carlo (MC; \cite{DUANE1987}) implemented within MATCH. These uncertainties scale with the number of stars in each cell, where more stars in a cell result in lower random uncertainties. From each cell's CMD, we generate 10,000 possible SFHs. We then calculate the 1-sigma error by calculating the 68th-percentile of the samples for the cell.  

\section{Results} \label{sec:results}
In Table \ref{tab:resolvedSFR}, we present the best fit star formation rates per time bin, number of stars in the cell (N), area in arcsec$^2$, mean age of the population, A$_V$, and dA$_V$, along with their cell indices and vertices. The mean age of the population in each cell is only included for a convenient reference. Some cells can be have a bimodal or trimodal age distribution, so please use this mean age with caution. Some numbers in this table have been rounded to save space, but the full machine readable table for all 2658 cells contain the measurements with full precision. 

\subsection{Star Formation Rate and Mass Maps} \label{ssec:SFR_maps}
\begin{figure*}[h]
    \centering
    \includegraphics[width=\textwidth]{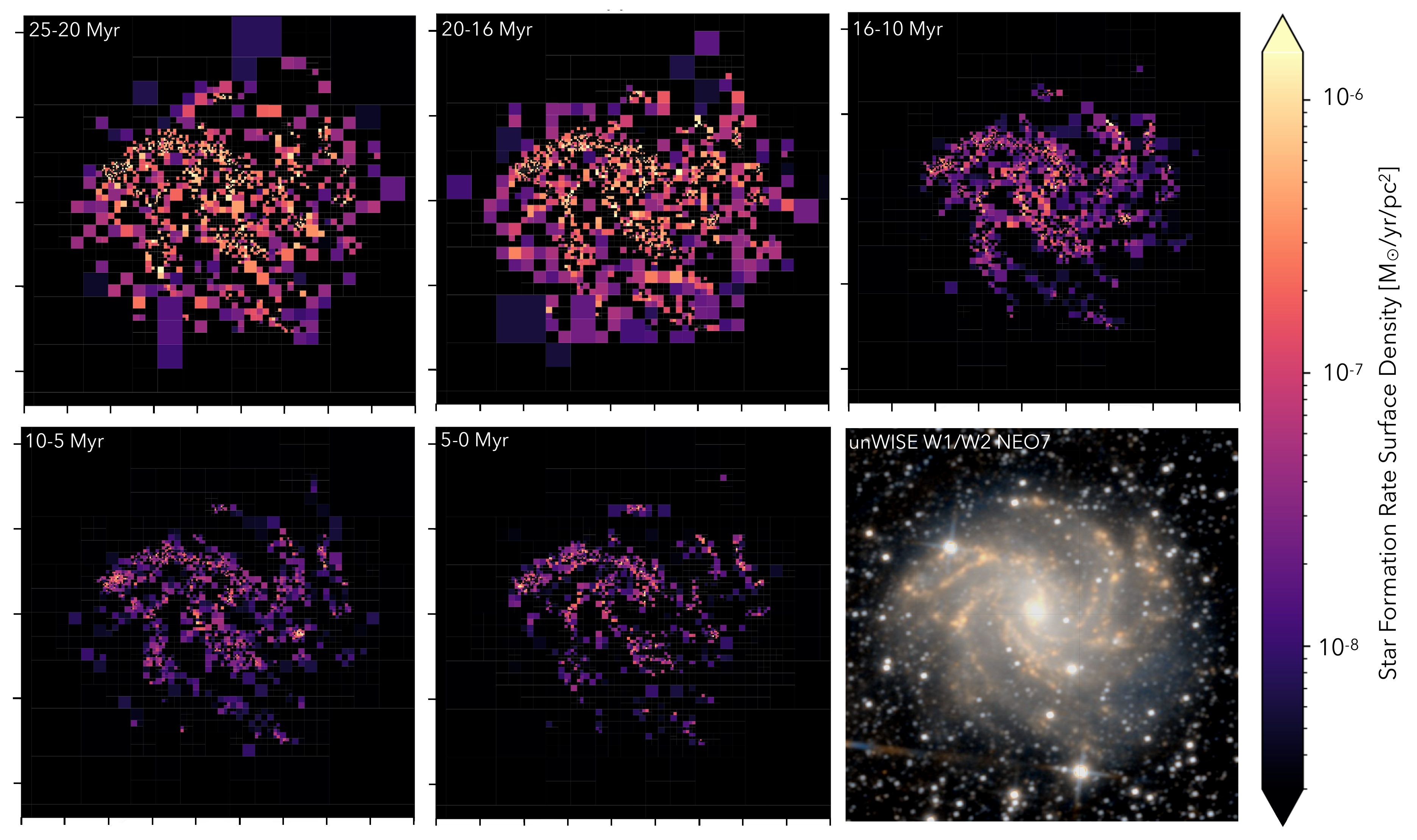}
    \caption{Maps of the spatially-resolved star formation rate as a function lookback time with linear time bins (RA=[60.03-60.226],dec=[308.50,308.95]). Each subfigure has the same dimensions, tick marks, and extent as Figure 2. We include a color image from the unWISE catalog \citep{Meisner2022} in W1 and W2 filters from NEO7, taken from legacysurvey.org.}
    \label{fig:sfh-linear}
\end{figure*}

\begin{figure*}[h]
    \centering
    \includegraphics[width=\textwidth]{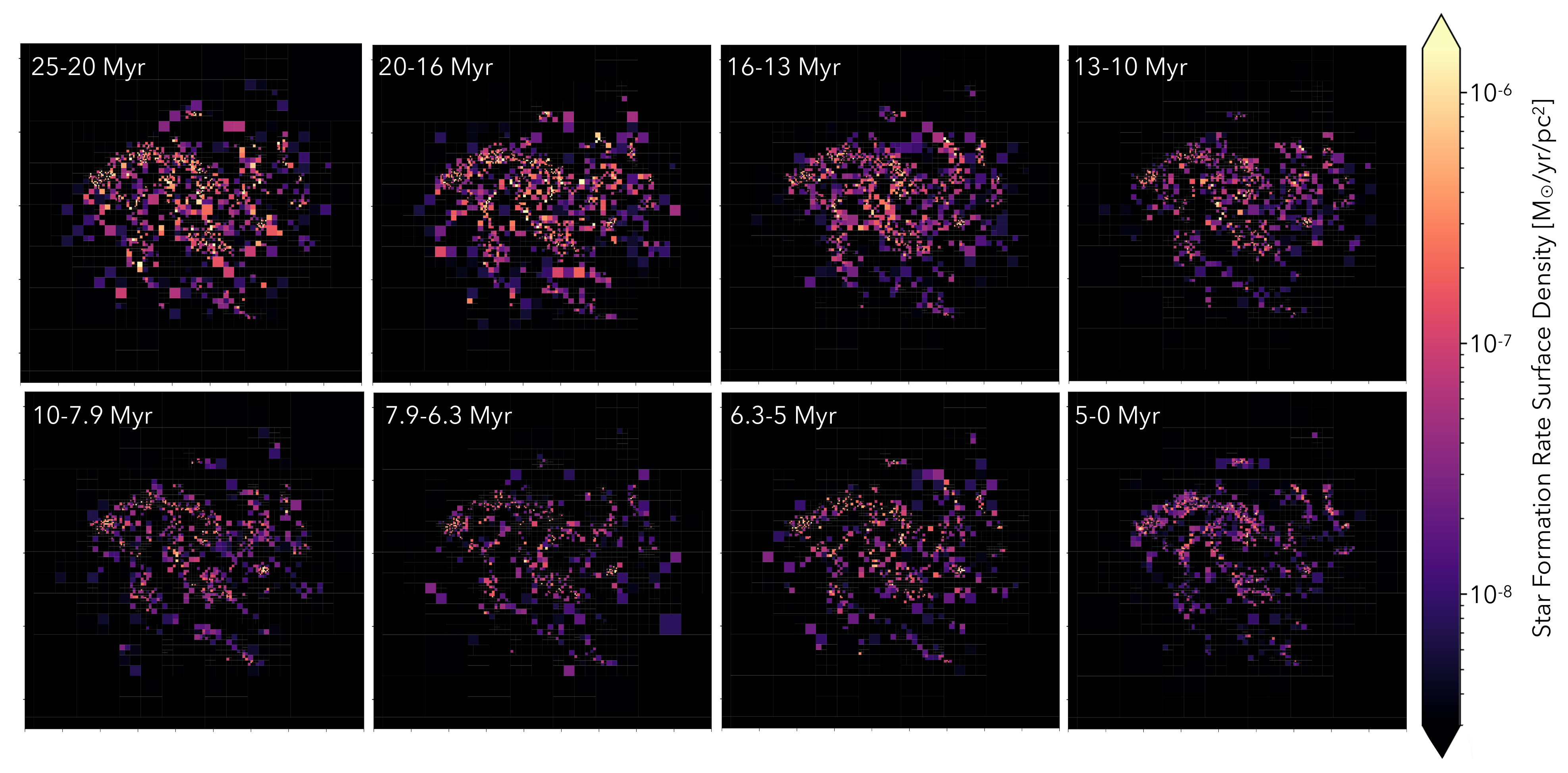}
    \caption{Maps of the spatially-resolved star formation rate as a function lookback time with logarithm time bins (RA=[60.03-60.226],dec=[308.50,308.95]). These plots use the same data as Figure \ref{fig:sfh-linear}, but use logarithm time bins to produce better time resolution at recent times. Each subfigure has the same dimensions, tick marks, and extent as Figure 2.}
    \label{fig:sfh-log}
\end{figure*}

In Figure \ref{fig:sfh-linear}, we present maps of the spatially-resolved star formation rate for NGC 6946 in linear time bins, and include a color image from the unWISE catalog \citep{Meisner2022}) in W1 and W2 filters to illustrate that the star formation in the youngest ages is mostly recovered despite the dust and we are not missing much embedded star formation. For every time bin, labeled in the upper left corner of each panel, we create maps with the best fit star formation rate setting the value of intensity of each pixel. These rates are then converted to star formation rate intensity by dividing by the area of the cell in corrected for the inclination of 32.8 degrees \citep{deBlok2008}. In Figure \ref{fig:sfh-log}, we present the spatially-resolved star formation history for NGC 6946 with log time bins for higher time resolution at younger ages.

\begin{deluxetable*}{cccccccccccccccccccccc}[h]
\tabletypesize{\tiny}
\rotate
\tablenum{3}
\tablecaption{Sample of SFRs over Time \label{tab:resolvedSFR}}
\tablewidth{0pt}
\setlength{\tabcolsep}{2pt}
\tablehead{
\colhead{i} & \colhead{RA-NE} & \colhead{Dec-NE} & \colhead{RA-NW} & \colhead{Dec-NW} &\colhead{RA-SE} & \colhead{Dec-SE} &\colhead{RA-SW} & \colhead{Dec-SW} &\colhead{N} & \colhead{Area} & \colhead{SFR 0-6.7} & \colhead{SFR 6.7-6.8} & \colhead{SFR 6.8-6.9} & \colhead{SFR 6.9-7.0} & \colhead{SFR 7.0-7.1} & \colhead{SFR 7.1-7.2} & \colhead{SFR 7.2-7.3} & \colhead{SFR 7.3-7.4} & \colhead{A$_V$} & \colhead{dA$_V$} & \colhead{Age} \\
\colhead{} & \colhead{$\arcdeg$} & \colhead{$\arcdeg$} & \colhead{$\arcdeg$} & \colhead{$\arcdeg$} &\colhead{$\arcdeg$} & \colhead{$\arcdeg$} &\colhead{$\arcdeg$} & \colhead{$\arcdeg$} & \colhead{} & \colhead{$\arcsec^2$} & \colhead{1e-3$M_\odot/yr$} & \colhead{1e-3$M_\odot/yr$} & \colhead{1e-3$M_\odot/yr$} & \colhead{1e-3$M_\odot/yr$} & \colhead{1e-3$M_\odot/yr$} & \colhead{1e-3$M_\odot/yr$} & \colhead{1e-3$M_\odot/yr$} & \colhead{1e-3$M_\odot/yr$} & \colhead{} & \colhead{} & \colhead{Myr}
}
\startdata
8 & 308.51220 & 60.150921 & 308.48380 & 60.150921 & 308.48380 & 60.136787 & 308.51220 & 60.136787 & 17 & 2588.9 & $0.00\substack{+0.14\\-0.00}$ & $0.00\substack{+0.57\\-0.00}$ & $0.00\substack{+0.64\\-0.00}$ & $0.06\substack{+0.62\\-0.06}$ & $1.19\substack{+0.12\\-1.04}$ & $0.00\substack{+0.64\\\-0.00}$ & $0.00\substack{+0.58\\\-0.00}$ & $0.00\substack{+0.90\\\-0.00}$ & 0.80 & 0.00 & 11.1\\
9 & 308.51220 & 60.165054 & 308.48380 & 60.165054 & 308.48380 & 60.150921 & 308.51220 & 60.150921 & 6 & 2588.9 & $0.00\substack{+0.10\\-0.00}$ & $0.00\substack{+0.41\\-0.00}$ & $0.00\substack{+0.36\\-0.00}$ & $0.00\substack{+0.38\\-0.00}$ & $0.00\substack{+0.43\\-0.00}$ & $0.00\substack{+0.42\\-0.00}$ & $1.04\substack{+0.26\\-0.83}$ & $0.00\substack{+0.86\\\-0.00}$ & 0.95 & 0.05 & 17.8\\
12 & 308.59740 & 60.263989 & 308.48380 & 60.263989 & 308.48380 & 60.207455 & 308.59740 & 60.207455 & 18 & 41422.6 & $0.00\substack{+0.22\\-0.00}$ & $0.00\substack{+1.11\\-0.00}$ & $0.00\substack{+1.22\\-0.00}$ & $2.01\substack{+0.00\\-1.98}$ & $0.00\substack{+1.22\\-0.00}$ & $0.00\substack{+1.20\\-0.00}$ & $0.00\substack{+2.76\\\-0.00}$ & $5.80\substack{+0.21\\\-5.80}$ & 0.95 & 0.30 & 18.9\\
14 & 308.54060 & 60.136787 & 308.51220 & 60.136787 & 308.51220 & 60.122654 & 308.54060 & 60.122654 & 7 & 2588.9 & $0.23\substack{+0.10\\-0.20}$ & $0.00\substack{+0.78\\-0.00}$ & $0.00\substack{+0.64\\-0.00}$ & $0.00\substack{+0.78\\-0.00}$ & $0.00\substack{+0.82\\-0.00}$ & $0.00\substack{+0.09\\-0.00}$ & $0.00\substack{+1.88\\\-0.00}$ & $0.00\substack{+4.20\\\-0.00}$ & 1.35 & 0.00 & 4.4\\
15 & 308.54060 & 60.150921 & 308.51220 & 60.150921 & 308.51220 & 60.136787 & 308.54060 & 60.136787 & 60 & 2588.9 & $4.29\substack{+0.81\\-1.76}$ & $0.00\substack{+7.40\\-0.00}$ & $0.00\substack{+5.96\\-0.00}$ & $10.0\substack{+0.07\\-9.62}$ & $0.00\substack{+8.68\\-0.00}$ & $0.00\substack{+13.9\\-0.00}$ & $17.5\substack{+3.98\\\-17.5}$ & $0.00\substack{+31.6\\-0.00}$ & 1.20 & 1.05 & 13.2\\
16 & 308.54060 & 60.165054 & 308.51220 & 60.165054 & 308.51220 & 60.150921 & 308.54060 & 60.150921 & 54 & 2588.9 & $1.95\substack{+0.27\\-1.05}$ & $0.00\substack{+3.11\\-0.00}$ & $0.00\substack{+3.32\\-0.00}$ & $2.42\substack{+1.80\\-2.42}$ & $0.00\substack{+2.81\\-0.00}$ & $3.37\substack{+2.82\\-2.83}$ & $0.00\substack{+5.23\\\-0.00}$ & $11.9\substack{+10.1\\\-8.37}$ & 1.40 & 0.00 & 17.5\\
17 & 308.54060 & 60.179188 & 308.51220 & 60.179188 & 308.51220 & 60.165054 & 308.54060 & 60.165054 & 40 & 2588.9 & $1.15\substack{+0.69\\-0.67}$ & $0.00\substack{+4.72\\-0.00}$ & $7.30\substack{+2.64\\-4.74}$ & $0.00\substack{+5.01\\-0.00}$ & $0.00\substack{+6.28\\-0.00}$ & $13.4\substack{+0.00\\-11.7}$ & $0.00\substack{+15.6\\\-0.00}$ & $0.00\substack{+46.4\\\-0.00}$ & 1.50 & 0.05 & 11.3\\
18 & 308.56900 & 60.108520 & 308.54060 & 60.108520 & 308.54060 & 60.094387 & 308.56900 & 60.094387 & 41 & 2588.9 & $0.00\substack{+0.91\\-0.00}$ & $0.00\substack{+14.1\\-0.00}$ & $57.3\substack{+0.21\\-27.7}$ & $0.00\substack{+16.6\\-0.00}$ & $0.00\substack{+13.3\\-0.00}$ & $0.00\substack{+11.0\\-0.00}$ & $0.00\substack{+16.7\\\-0.00}$ & $0.00\substack{+37.0\\\-0.00}$ & 1.50 & 1.50 & 7.1\\
19 & 308.55480 & 60.115587 & 308.54060 & 60.115587 & 308.54060 & 60.108520 & 308.55480 & 60.108520 & 3 & 647.2 & $0.12\substack{+0.11\\-0.12}$ & $0.00\substack{+0.76\\-0.00}$ & $0.00\substack{+0.67\\-0.00}$ & $0.00\substack{+0.62\\-0.00}$ & $0.00\substack{+0.53\\-0.00}$ & $0.00\substack{+0.82\\-0.00}$ & $1.24\substack{+2.02\\\-0.92}$ & $0.00\substack{+2.81\\\-0.00}$ & 1.40 & 0.00 & 16.6\\
20 & 308.55480 & 60.122654 & 308.54060 & 60.122654 & 308.54060 & 60.115587 & 308.55480 & 60.115587 & 6 & 647.2 & $0.00\substack{+0.24\\-0.00}$ & $0.00\substack{+1.21\\-0.00}$ & $2.25\substack{+0.52\\-1.83}$ & $0.00\substack{+1.29\\-0.00}$ & $0.00\substack{+1.17\\-0.00}$ & $0.00\substack{+1.23\\-0.00}$ & $0.00\substack{+1.96\\\-0.00}$ & $0.00\substack{+4.79\\\-0.00}$ & 1.45 & 0.00 & 7.1
\enddata
\tablecomments{Note: Some of the values have been rounded to save space. The machine readable table provided will have the full precision. Area listed is not corrected for inclination. i= index; N = number of stars}
\end{deluxetable*}

\clearpage

\begin{figure*}
    \centering
    \includegraphics[width=\textwidth]{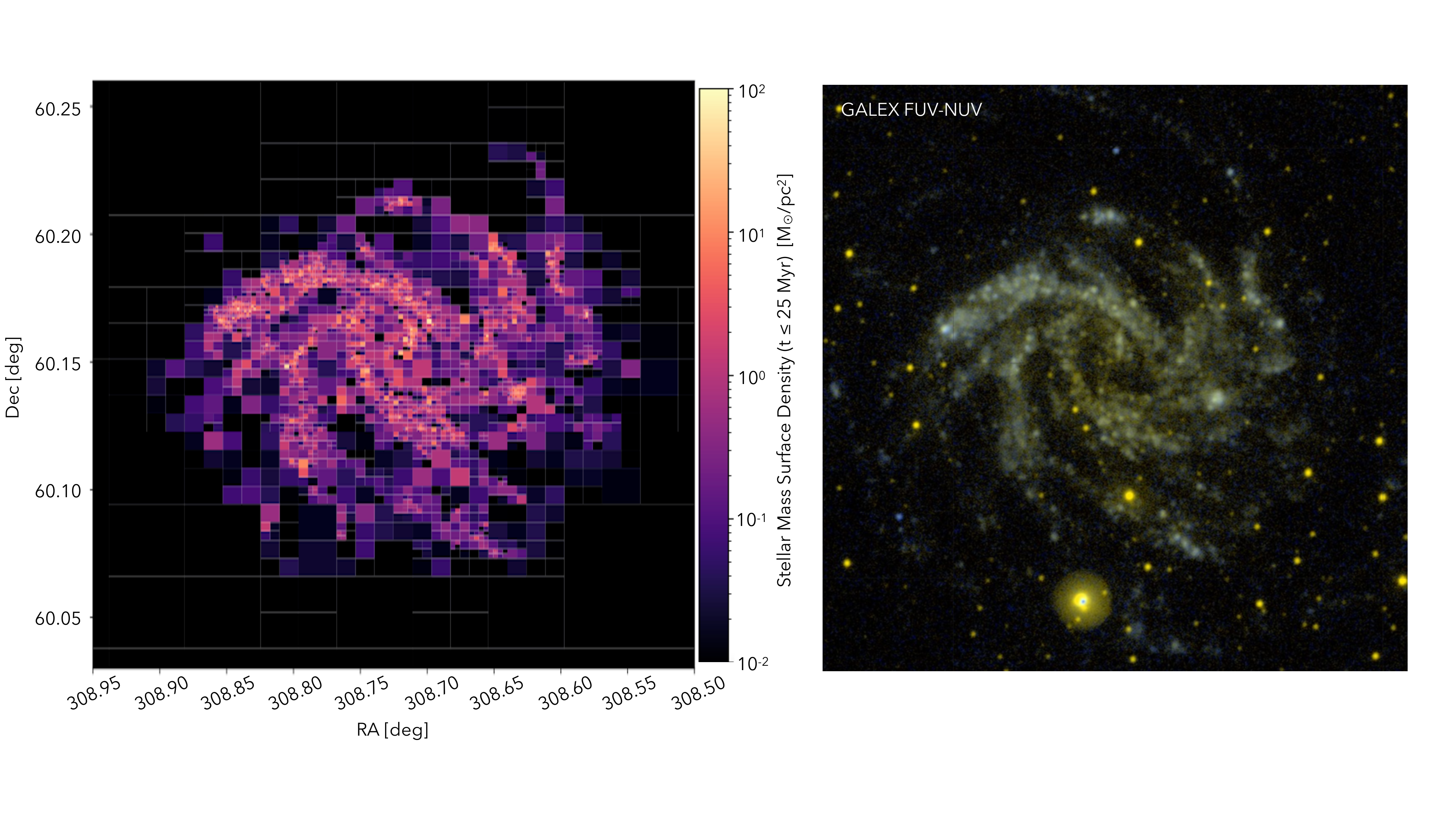}
    \caption{Left: Total stellar mass formed per unit area in the past 25 Myr. Right: GALEX NUV-FUV color image taken from legacysurvey.org. }
    \label{fig:mass_map}
\end{figure*}

We integrate the time bins to calculate the total mass formed over the last 25 Myr. In the left panel of Figure \ref{fig:mass_map}, we show the resulting mass surface density map. A majority of the recently formed mass is in the spiral arms, though there is a significant population of young stars outside of the spiral arms. The spatial distribution of the mass formed traces the resolved UV photometry of the galaxy fairly well. Our data and mass map look far more extended, particularly in the northwest and southeast arms, than the GALEX color image (right panel of Figure \ref{fig:mass_map}) due to the increased sensitivity of our data (Figure \ref{fig:footprint}). Our methods seem to be more sensitive to older star formation than that of GALEX (which would probe $<$10Myr), as these features appear most prominently in the 16-20 Myr time bins of Figures \ref{fig:sfh-linear} and \ref{fig:sfh-log}.

\subsection{Extinction Maps} \label{ssec:Extinction_maps}
\begin{figure*}
    \centering
    \includegraphics[width=\textwidth]{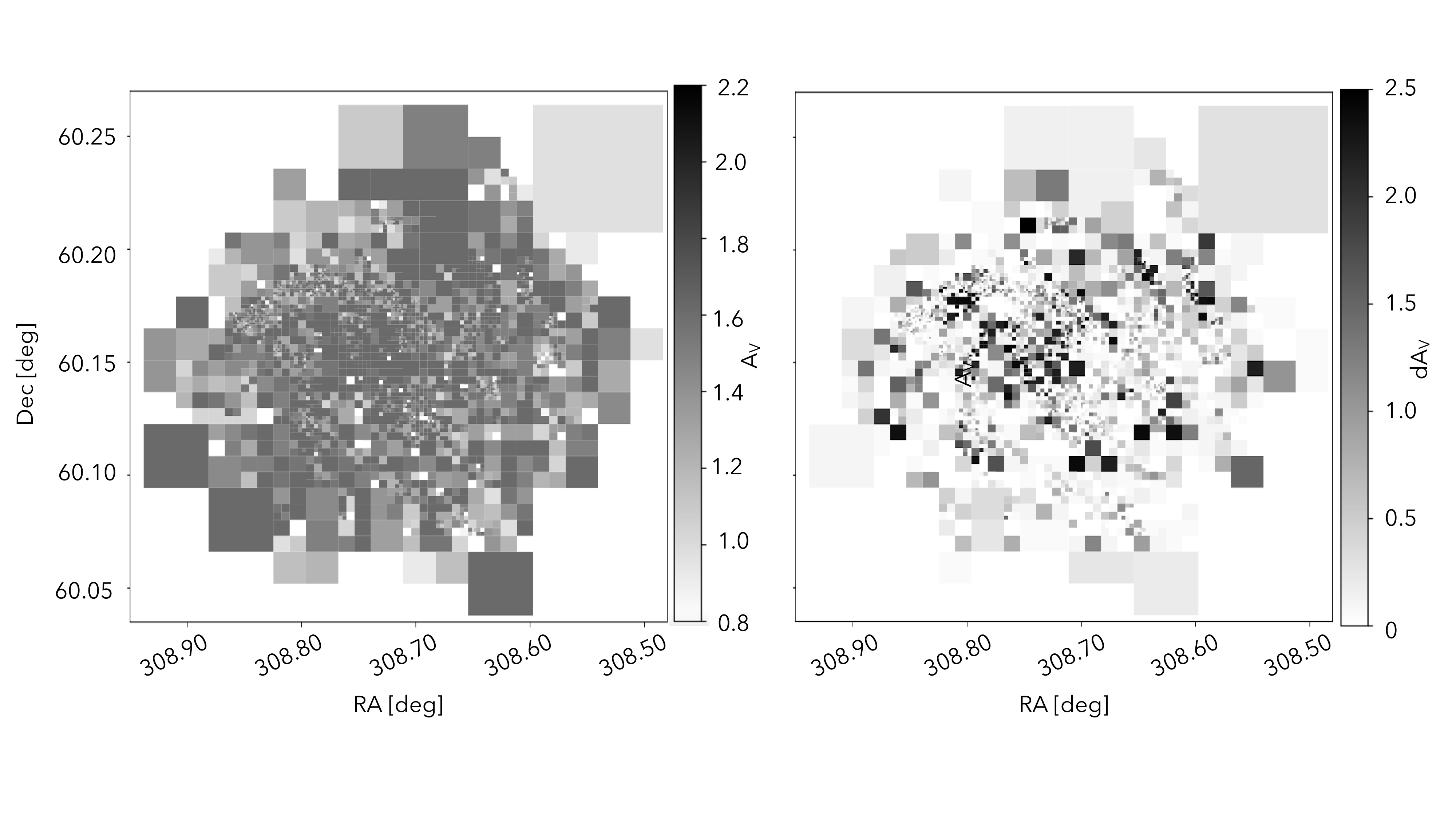}
    \caption{Left: Best fit foreground extinction A$_V$ of NGC 6946. Right: Best fit differential extinction dA$_V$ of NGC 6946.}
    \label{fig:AvdAv}
\end{figure*}

We recovered foreground extinction fairly uniform with mean of 1.4 and standard deviation of 0.3 (Figure \ref{fig:AvdAv}, left panel). This is slightly higher than the extinction of $A_V$=0.938 from \cite{Schlafly2011}. We present the differential extinction map in the right panel of Figure \ref{fig:AvdAv}. The areas of high differential extinction are in the spiral arms and appear to be very clumpy. Approximately 17\% of the cells have high measured differential extinction (dA$_V>$1), which could mean that we cannot detect some of the older stars in those grids. For more detail, see Section \ref{ssec:old-reliability}. The measured differential extinction show no correlation with ages measured in Section \ref{ssec:density_v_age}.

\subsection{Global SFH} \label{ssec:Total_SFR}
\begin{figure*}
    \centering
    \includegraphics[width=.6\textwidth]{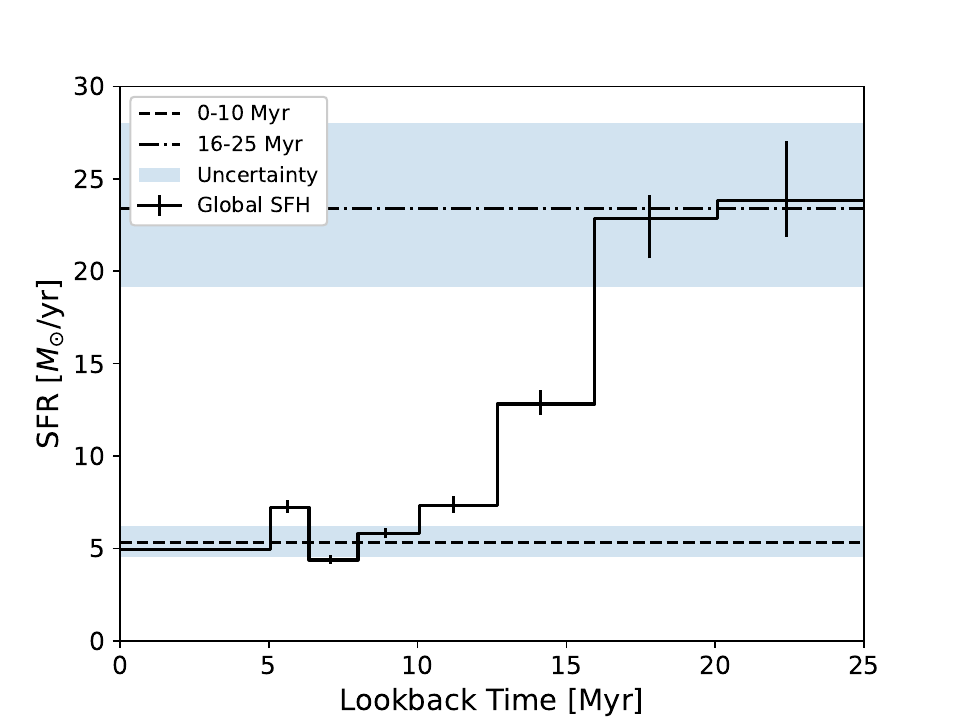}
    \caption{Coadded star formation rates across all spatial bins as a function of lookback time. The global star formation rate is much higher 16 Myr ago and has decreased in to be a more steady rate in the past 10 Myr.}
    \label{fig:global_sfh}
\end{figure*}

To derive a global star formation history for the galaxy, at each time bin, we integrate the SFR over all cells. We calculate the uncertainties due to the number of stars in the cells by adding the uncertainties of each spatial bin in quadrature. Then we bootstrap the uncertainties across spatial bins by sampling the number of cells 10,000 times with replacement to account for uncertainties due to binning the stars into cells. We then add the uncertainties obtained via bootstrapping to the random uncertainties in quadrature. We present these global star formation rates in Table \ref{tab:sfr} and plot them in Figure \ref{fig:global_sfh}. 

To obtain the star formation rates in the past 10 Myr, we integrated the star formation rates over time to obtain total mass formed in the past 10 Myr, then divided that total mass by 10 Myr. We calculated the uncertainties by adding the uncertainties of each time bin in quadrature. We find the global star formation rate over the past 10 Myr to be roughly constant at $5.31\substack{+0.88\\-0.78} M_\odot/$yr, shown in Figure \ref{fig:global_sfh}. We did the same for the global star formation rate 16-25 Myr ago, obtaining an SFR of $23.38\substack{+4.65\\-4.25} M_\odot/$yr. The SFR 16-25 Myr ago was roughly five times larger than the current ($\leq$ 10 Myr) star formation rate, with a monotonically decreasing SFR in the 6 Myr in between the two epochs. 

\subsection{Reliability of the Younger Time Bins} \label{ssec:young-reliability}
As with SFR measurement techniques that rely on UV data, dust is a big challenge to measuring young star formation. Stars are born from giant molecular clouds and are obscured by dust until a massive star forms, ionizes its birth cloud, clears out the material \citep{Lada2003}. This makes young stars incredibly challenging to observe. Our technique requires that the young stars are observable to measure star formation. The uncertainties due to dust for SFRs older than 8 Myr decrease significantly. Young stellar clusters emerge from the giant molecular clouds on timescales of 8 Myr, with very few stars remaining embedded after that \citep{Corbelli2017}. Additionally, upon visual inspection of WISE and GALEX images (Figures \ref{fig:sfh-linear} and \ref{fig:mass_map}, respectively), the star formation in the 0-5 Myr time bin seems well recovered. This suggests that it is unlikely that a significant fraction of the star formation is being missed, though additional measurements of SFR using infrared are necessary better constrain the impacts of dust.

We check the impact of the blends on the reliability of the SFR in the young time bins. Our choice to include the 16 sources flagged as blends in 14 cells had been made to include as many young stars as possible in our MATCH fits. We ran MATCH again on two of these cells removing the blends from the observed CMD. We find that there is no impact on the measured SFH of these two cells. However, without the blends, the uncertainty of the SFR in the youngest time bin decreased by two orders of magnitude. The inclusion of the blends in our CMD-fitting gives us more conservative uncertainties on the SFRs of the younger time bins.

Additionally, we compare the SFR measurements to that from the literature. The current SFR is consistent with SFRs measured via methods probing the youngest stars. Previous measurements of the SFR within the last 5 Myr from H$\alpha$ measurements find an SFR of $\sim$ 4 $M_\odot/$yr (no reported uncertainties, \cite{Sauty1998}) and 5.7 $\pm$ 1.7 $M_\odot/$yr \cite{Botticella2012}, which both agree with our measurement of $4.93 \substack{+0.22 \\ -0.23}$  (Table 4, row 1) within uncertainties. \cite{Kennicutt2011} measured a $\lesssim$ 5 Myr SFR of $\simeq$ 7.1 $M_\odot/$yr with $H\alpha$ and 24 $\mu$m observations, with no uncertainties reported. This higher star formation rate is more consistent with our SFR in the 5-6.3 Myr time bin of $7.21 \substack{+0.58 \\ -0.52}$ ((Table 4, row 2). Measurements of the SFR obtained with far-ultraviolet (FUV) tend to probe timescales over the past 10 Myr. \cite{Botticella2012} measures a FUV SFR 9.1 $\pm$ 2.7 $M_\odot/$yr, which is more consistent with our measurements of the SFRs 10-16 Myr ago, $7.33 \substack{+0.70 \\ -0.66}$ and $12.81 \substack{+1.04 \\ -0.95}$ (Table 4, rows 5-6).

\subsection{Reliability of Older Time Bins} \label{ssec:old-reliability}
Another challenge of utilizing UV data to measure star formation rates is that UV primarily probes young star formation, as seen in Figure \ref{fig:cmd} where the log(age)=7.4 isochrone lies very close to the completeness limit of our data. This limitation is reflected in the high uncertainties (at least an order of magnitude higher than the rest of the time bins) in the star formation rate in the oldest time bin, in Table \ref{tab:sfr}. Thus we exclude the log(age)=7.4-7.5 time bin from the our analysis in the paper. In addition, we perform several tests to check the reliability of the SFR in the two oldest time bins (log(age)=7.2-7.3 and log(age)=7.3-7.4) by simulating model CMDs from the resulting SFRs. 

We perform tests to check the impact of our chosen completeness limits on our measured star formation rates in the two oldest time bins. First, we check that the model CMDs created accurately model the observed CMDs for a selection of cells at varying stellar densities. We create these model CMDs by using the same parameters (i.e. completeness limits, binary fraction) we used to fit the star formation histories, as well as the output best fit metallicities and star formation rates from our results. We then check that the model has enough stars in the oldest two time bins log(age)=7.2-7.3 and log(age)=7.3-7.4 to allow for good measurements. 

We also test for density-dependent effects which might arise due to the brighter completeness limit in the highest density regimes. We want to ensure that the measured SFRs of the oldest two time bins is consistent with the number of stars in their observed CMDs and reflect an accurate measurement. We check this by performing two tests. First, to measure the minimum expected percentage of older stars, we simulate constant star formation histories at different constant SFRs to measure the percentage of older (log(age)=7.2-7.4) stars out of the total (log(age)=6.6-7.4) at varying stellar densities. To check the impact of the input SFR on the percentage of older stars, we choose an SFR of 13.17 $M_\odot/$yr (the average global SFR), 0.1 $M_\odot/$yr (the highest measured SFR of a cell), and 0.005 $M_\odot/$yr (an average SFR value in a cell). We present the resulting systematics from these tests below (Table \ref{tab:reliability}). There is some stochasticity in the percentage of modeled older stars for a constant SFH at SFR=0.005 $M_\odot/yr$ likely due to the small numbers of stars in the model. Second, we measure the percentage of observed stars that fell in the two oldest time bins for varying stellar densities in a selection of spatial cells that have a spike in SFR in the older time bins. For the highest stellar densities ($\geq$ 12 stars/arcsec$^2$), the log(age)=7.2-7.3 and 7.3-7.4 time bins contain 20 $\pm$ 10\% of the total number of stars, whereas the lower stellar density bins have 26 $\pm$ 10\% of the total number of stars. Fortunately, there are only 5 cells that have these high densities, which is not significant enough to impact the high measured global SFR in the older time bins. The lower density bins have a sufficient percentage of stars above the minimum expected percentage to have the accurate SFRs measured in the oldest time bins, appearing to be less impacted by the completeness limit concerns than the 5 cells in the high density regime.

Finally, we compare the SFR measured in the oldest time bins to that obtained via supernova rates, which probes older star formation, from 30 Myr (assuming single star evolution) to 100 Myr (assuming binary star evolution; \cite{Jennings2014,Smartt2009,Zapartas2021}) and requires an SFR of at least 12.1 $\pm$ 3.7 $M_\odot/yr$ \cite{Eldridge2019}. We convert the measured SFR, $\psi$(t), of the two oldest time bins to an estimated core-collapse supernova rate, R(t)$_{CC}$. We utilize the formalism from \cite{Blanc2008}, using the same assumptions (all stars in the suitable mass range $m_u^{CC}-m_l^{CC}$ supernova, the number fraction of stars that supernova in the time range and number of stars per unit mass of the stellar generation do not vary with time, since the time range we are looking at is very short) and resulting equation (Equations \ref{eq:conversion} and \ref{eq:K}) from \cite{Botticella2012}. 

\begin{equation} \label{eq:conversion}
    R(t)_{CC} = K_{CC} \times \psi(t)
\end{equation}
where
\begin{equation} \label{eq:K}
    K_{CC} = \frac{\int_{m_l^{CC}}^{m_u^{CC}} \phi(m) \,dm}{\int_{m_l}^{m_u} m\phi(m) \,dm}
\end{equation}

For consistency, we choose the IMF, $\phi(m) \propto m^{-\alpha_i}$, to be the Kroupa IMF ($\alpha_0 = +0.3,\ 0.01 \leq m/M_\odot < 0.08; \quad \alpha_1 = +1.3,\ 0.08 \leq m/M_\odot < 0.50; \quad \alpha_2 = +2.3,\ 0.50 \leq m/M_\odot$), \cite{Kroupa2001}). We utilize the minimum progenitor masses to supernova for log(age)=7.2-7.3 to be 11.73 M$_\odot$ and log(age)=7.3-7.4 for 10.28 M$_\odot$ from the Padova stellar evolutionary models \citep{Marigo2008} used in the calculation of the SFH.

For log(age)=7.2-7.3, we derive a core-collapse supernova rate of $0.15 \substack{+0.02\\-0.02}$ SN/yr. For log(age)=7.3-7.4, a ccSN rate = $0.19\substack{+0.03\\-0.02}$ SN/yr, with uncertainties only propagated from SFR in the two time bins. This doesn't account for the systematic uncertainties due to the IMF and estimated progenitor mass. We estimate the uncertainty due to choice of IMF by comparing the ccSN rates obtained with the high-mass IMF from \cite{Weisz2015}. The ccSN rate obtained for log(age)=7.2-7.3 is $0.09\substack{+0.01\\-0.01}$ SN/yr and $0.12\substack{+0.02\\-0.02}$ SN/yr for log(age)=7.3-7.4, which is more aligned with the observed supernova rate of 0.1 SN/yr \citep{Eldridge2019}. A 6\% change in the slope of the IMF changes the ccSN rate by 60\%, dominating the uncertainty of this calculation. We estimate the uncertainty due to progenitor mass by selecting progenitor masses from two additional stellar evolutionary models, PARSEC \citep{Bressan2012} and Geneva \citep{Ekstrom2012}, and taking the standard deviation the ccSN rate obtained with the three different models. The uncertainty due to choice of progenitor mass is 0.7\% and 4\% for log(age)=7.2-7.3 and log(age)=7.3-7.4, respectively.

\section{Discussion} \label{sec:discussion}
In Section \ref{ssec:local}, we analyze two regions with high recent star formation and present their star formation histories. Finally, in Section \ref{ssec:density_v_age}, we examine the relationship between stellar density and age.

\begin{deluxetable*}{cccccccc}[h]
\tablenum{5}
\tablecaption{Percentage of Modeled Older Stars for Simulated Constant SFH at SFR of 13.17 $M_\odot/$yr (the average global SFR), 0.1 $M_\odot/$yr (the highest measured SFR of a cell), and 0.005 $M_\odot/$yr (an average SFR value in a cell)} \label{tab:reliability}
\tabletypesize{\normalsize}
\tablewidth{45pt}
\tablehead{
\colhead{Density Bin} & \colhead{SFR = 13.17} & \colhead{SFR = 0.1} & \colhead{SFR = 0.005}\\
\colhead{[stars/arcsec$^2$]} & \colhead{$M_\odot/yr$} & \colhead{$M_\odot/yr$} & \colhead{$M_\odot/yr$}
}
\startdata
0-2 & 18\% & 18\% & 25\%\\
2-4 & 18\% & 19\% & 15\%\\
4-6 & 18\% & 18\% & 14\%\\
6-8 & 17\% & 18\% & 15\%\\ 
8-10 & 17\% & 17\% & 20\%\\  
10-11 & 17\% & 16\% & 10\%\\ 
11-12 & 18\% & 17\% & 22\%\\ 
12-13 & 15\% & 11\% & 3\%\\
13-18 & 14\% & 11\% & 19\%\\ 
\enddata
\end{deluxetable*}

\begin{deluxetable*}{ccc}[h]
\tablenum{4}
\tablecaption{Global SFR over Time \label{tab:sfr}}
\tabletypesize{\normalsize}
\tablewidth{45pt}
\tablehead{
\colhead{Time Bin} &
\colhead{\hspace{1cm}Time Bin} & \colhead{\hspace{1cm}SFR\hspace{1cm}}\\ \colhead{[Myr]} &
\colhead{\hspace{1cm}[log(yr)]} & \colhead{\hspace{1cm}[$M_\odot/yr$]\hspace{1cm}} 
}
\startdata
4-5 & \hspace{1cm}6.6-6.7 & \hspace{1cm}$4.93 \substack{+0.22 \\ -0.23}$\\
5-6.3 & \hspace{1cm}6.7-6.8 & \hspace{1cm}$7.21 \substack{+0.58 \\ -0.52}$\\
6.3-8 & \hspace{1cm}6.8-6.9 & \hspace{1cm}$4.37 \substack{+0.42 \\ -0.36}$\\
8-10 & \hspace{1cm}6.9-7.0 & \hspace{1cm}$5.80 \substack{+0.45 \\ -0.40}$\\
10-12.5 & \hspace{1cm}7.0-7.1 & \hspace{1cm}$7.33 \substack{+0.70 \\ -0.66}$\\
12.5-16 & \hspace{1cm}7.1-7.2 & \hspace{1cm}$12.81 \substack{+1.04 \\ -0.95}$\\
16-20 & \hspace{1cm}7.2-7.3 & \hspace{1cm}$22.84 \substack{+2.67 \\ -3.18}$\\
20-25 & \hspace{1cm}7.3-7.4 & \hspace{1cm}$23.82 \substack{+3.81 \\ -2.81}$\\
\enddata
\end{deluxetable*}

\subsection{Local Star Formation History} \label{ssec:local}
We analyze two regions with high star-forming activity, the Hodge Complex \citep{Hodge1967} roughly centered at 20:34:34.80 +60:08:18.60 and the HII region at the tip of the northeast spiral arm roughly centered at 20:35:22.57 +60:10:14.70. The cells of the regions of interest used in this analysis are flagged in the machine readable table. 

\begin{figure*}
    \centering
    \includegraphics[width=\textwidth]{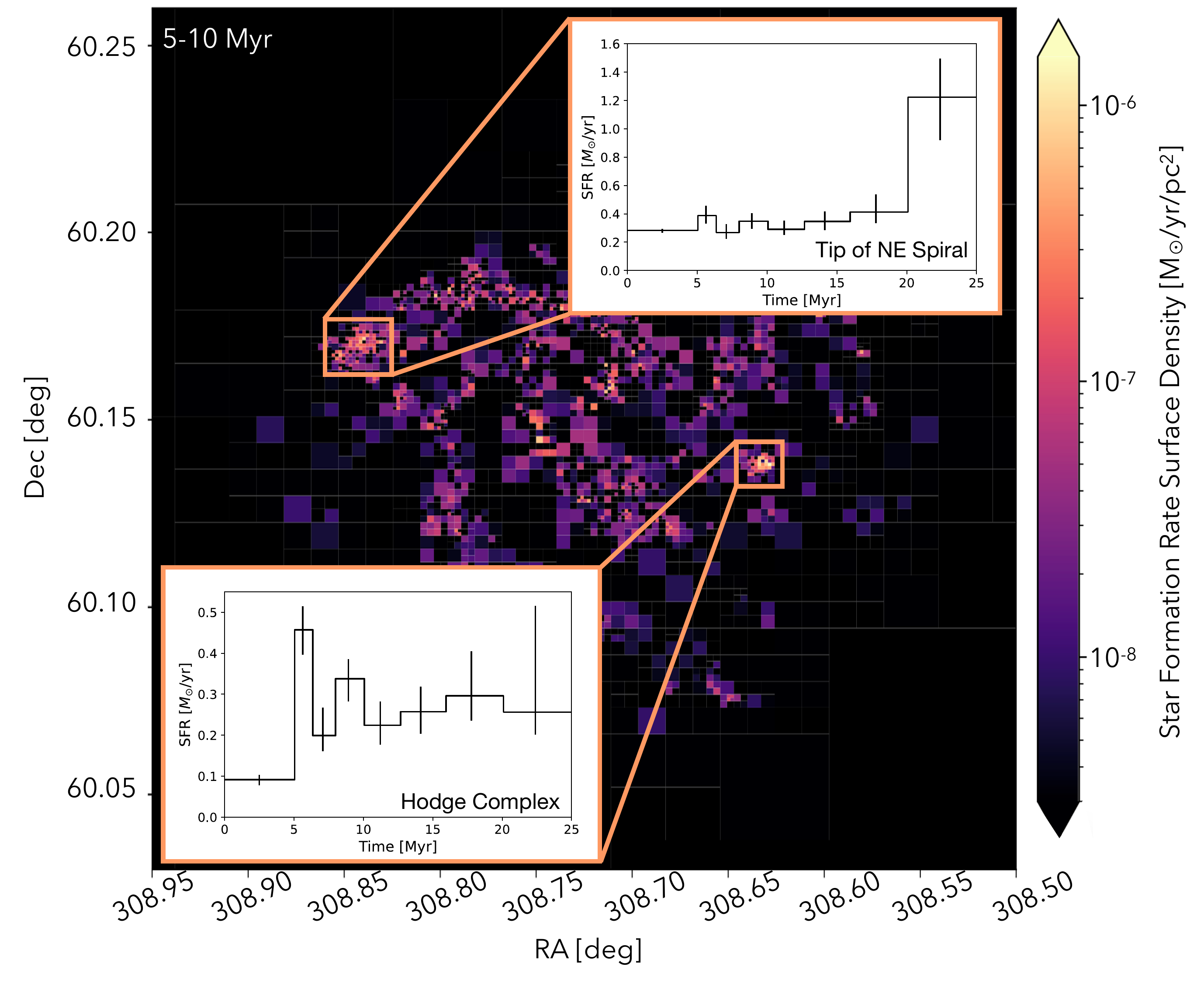}
    \caption{Coadded star formation histories of two regions of interest, the large HII region in the northeast spiral arm and Hodge Complex, with the approximate locations of the regions used in the local star formation history. Exact indices of the regions used in the star formation history calculations are flagged in the machine readable table.}
    \label{fig:region_sfh}
\end{figure*}

To obtain the star formation histories of these regions, we sum the star formation rates of each cell in the region per time bin. We then add the random uncertainties in quadrature. The locations of the regions and their star formation histories are presented in Figure \ref{fig:region_sfh}.

The Hodge Complex is a super star cluster containing multiple young star clusters and has been extensively studied due to the high concentration of star formation \cite{Gil2009,Efremov2007,Efremov2016}. We present the star formation history of this region in the lower left corner of Figure \ref{fig:region_sfh}. This region appears to have constant star formation over the past 6.3-25 Myr with a peak in star formation in around 5-6.3 Myr and drop in star formation in the most recent 5 Myr. Interestingly, despite being a mere 1295 square arcseconds, which is 0.05\% of the total size of our coverage area, this tiny region contains 3\% of the total mass formed up to 6.3 Myr and 1.8\% of the total mass formed up in the past 25 Myr. This region has had a more recent star formation episode than seen for the globally decreasing star formation rate. 

Another region of interest is the HII region in tip of the northeast spiral arm. Unlike the Hodge Complex, the star formation history of this region roughly follows the star formation history of galaxy. Similar to the Hodge Complex, this small region of the galaxy contains a significant portion of the recent star formation in the galaxy. Despite it being 0.05\% of our total coverage (roughly 1334 square arcseconds), this region contains 5.6\% of total mass of NGC 6946 formed in the last 6.3 Myr and 3.9\% over the last 25 Myr. There is a large peak in older star formation 20-25 Myr ago relative to the flatter SFH in the past 20 Myr. Since the global SFH has a more gradual decrease in SFR over time, we check that this peak in SFR is not due to a systematic related to completeness. We check SFRs of the locations with the highest stellar density and find 1.5\% of SFR in oldest time is attributed to those high density regions, making an insignificant contribution to the high SFR.

Ultimately, these regions of interest are only a small portion of the young star formation in NGC 6946. This points to overall star formation across the galaxy contributing to the peak in the global SFH.

\subsection{Density versus Age} \label{ssec:density_v_age}
Initially, we measured the characteristic age of the population in each cell by randomly sampling their star formation histories 50,000 times. We then obtain 16th, 50th, and 84th percentile time bins. Due to the the double peak distribution of many of the star formation histories, we found no correlation between stellar density and characteristic age for this population. 

\begin{figure*}
    \centering
    \includegraphics[width=.6\textwidth]{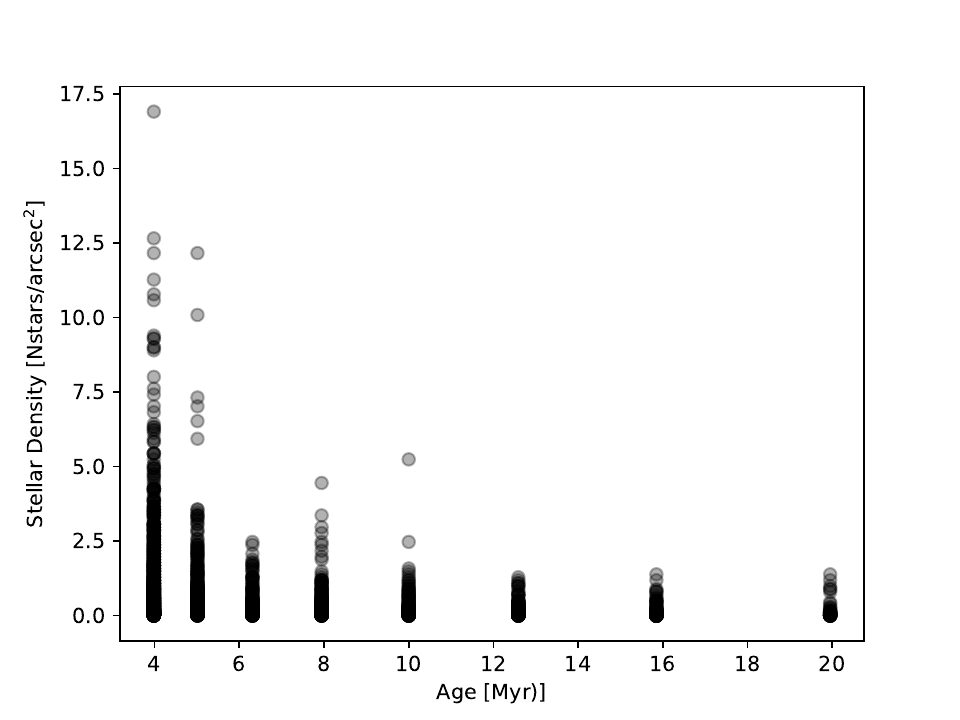}
    \caption{This plot shows the relationship between the age of the youngest detectable population, described in Section \ref{ssec:density_v_age}, and the stellar density, described in Section \ref{ssec:ASTs}, where each point represents a cell. There seems to be a correlation between stellar density and the youngest ages detected, where the younger the stars in the cell are, the more dense that cell is. This correlation seems to flatten after 12 Myr.}
    \label{fig:density-age}
\end{figure*}

Thus instead of looking at the characteristic age of the population, we find the youngest age bin in which we detect star formation in each cell, using the following criteria: the SFR of the age bin is at least 3\% of the total star formation rate, the age bin contains 1\% of the total mass formed, and and the lower bound SFR uncertainty of that time bin must be greater than or equal to zero. In Figure \ref{fig:density-age}, we see that the youngest detected population are denser, with the stellar density remaining between 0-2.5 stars/arsec$^2$ after 12 Myr. A possible explanation is that stars have migrated from their initial site of formation over 12 Myr to populate the field. This is consistent with the timeline of stars emerging for their giant molecular clouds by roughly 8 Myr (see Section \ref{ssec:young-reliability}) then populating the field over time. More work will need to be done to confirm that stellar migration is being observed. We plan to model this in our future work. 

\section{Conclusions}\label{sec:conclusions}
In this paper, we presented the spatially-resolved star formation history of NGC 6946 in the last 25 Myr, measured using resolved NUV stellar photometry. We implemented a quadtree algorithm to devise a spatial grid, and we measured the SFH independently for each cell using CMD-fitting. We summarize our main findings below.
\begin{itemize}
    \item We measure the global SFR over the last 25 Myr to be $13.16 \substack{+0.91\\-0.79} M_\odot/yr$. 
    \item 16-25 Myr ago, the SFR was $23.38\substack{+2.43\\-2.11} M_\odot/yr$. The SFR then monotonically decreases between 10-16 Myr, reaching a steady recent SFR in the past 10 Myr of $5.31\substack{+0.18\\-0.17} M_\odot/yr$.
    \item We present the star formation histories of the Hodge Complex and the HII region at the tip of northeast spiral arm. Both contain a higher amount of recent star formation than expected for regions of their size. The Hodge Complex shows more recent star formation relative to the declining global star formation.
\end{itemize}

\section{Acknowledgements}
We thank the reviewer for their helpful comments that improved the paper. This research is based on observations made with the NASA/ESA Hubble Space Telescope obtained from the Space Telescope Science Institute, which is operated by the Association of Universities for Research in Astronomy, Inc., under NASA contract NAS 5–26555. These observations are associated with program GO-15877.

\facilities{HST(WFC3/UVIS)}

\software{astropy \citep{astropy:2013,astropy:2018,astropy:2022}, DOLPHOT \citep{Dolphin2000,Dolphin2016}, GeoPandas \citep{geopandas2020}, \citep{Dolphin2002,dolphin2012,dolphin2013}, Matplotlib \citep{Hunter2007}, NumPy \citep{harris2020}, Pandas \citep{mckinney2010,pandas2020}.}

\clearpage

\bibliography{ngc6946}{}
\bibliographystyle{aasjournal}

\end{document}